# Determination of the scale of coarse graining in earthquake network

SUMIYOSHI ABE[1,2] and NORIKAZU SUZUKI[3]

[1] *Department of Physical Engineering, Mie University, Mie 514-8507, Japan*

[2] *Institut Supérieur des Matériaux et Mécaniques Avancés, 44 F. A. Bartholdi, 72000 Le Mans, France*

[3] *College of Science and Technology, Nihon University, Chiba 274-8501, Japan*

PACS   89.75.Da   —   Systems obeying scaling laws

PACS   91.30.–f   —   Seismology

PACS   05.65.+b   —   Self-organized systems

PACS   05.90.+m   —   Other topics in statistical physics, thermodynamics, and nonlinear dynamical systems

**Abstract.** — In a recent paper [ABE S. and SUZUKI N., *Europhys. Lett.*, 65 (2004) 581], the concept of earthquake network has been introduced in order to describe complexity of seismicity. There, the cell size, which is the scale of coarse graining needed for constructing an earthquake network, has remained as a free parameter. Here, a method is presented for determining it based on the scaling behavior of the network. Quite remarkably, both the exponent of the power-law connectivity distribution and the clustering coefficient are found to approach the respective universal values and remain invariant as the cell size becomes larger than a certain value, $l_*$, which depends on the number of events contained in the analysis, in general. This $l_*$ fixes the scale of coarse graining. Universality of the result is demonstrated for all of the networks constructed from the data independently taken from California, Japan and Iran.

Seismicity is a typical complex phenomenon. In spite of the long tradition of seismology, microscopic dynamics governing interaction/correlation between earthquakes is still largely unknown, although there are indications that the dynamics of a sequence of aftershocks seems to be of the glassy type [1], for example. It is empirically known that an earthquake can be triggered by a foregoing one more than 1000 km away [2]. In addition, both spatial distance [3] and time interval [4,5] between two successive events obey specific statistical laws, which are not Poissonian. These facts make it natural to put a working hypothesis that two successive events are indivisibly correlated at the statistical level, no matter how distant they are.

Recently, the concept of earthquake network has been introduced in Ref. [6]. The construction of an earthquake network proposed there is as follows. A geographical region under consideration is divided into a lot of small cubic cells, the linear dimension of each of which is denoted here by $L$. A cell is regarded as a vertex if earthquakes occurred therein. Two successive earthquakes define an edge between two vertices, which effectively replaces complex event-event correlation mentioned. If two successive events occur in the same cell, then they form a tadpole-loop (i.e., a self-loop or a bubble). This procedure allows one to map seismic data to a growing stochastic network unambiguously once the cells are set.

A couple of comments on the above construction are in order. Firstly, a full earthquake network is a directed one. Directedness is irrelevant to statistical analysis of connectivity since in-degree and out-degree are identical for each vertex except the first and last vertices in the analysis. Secondly, a full earthquake network should be reduced to an undirected simple network when small-worldness of the network is considered. There, tadpoles have to be removed and each multiple edge be replaced by a single edge

(see Fig. 1).

The earthquake networks have been constructed in this way from the real seismic data, and then it has been found [6,7] that they are scale-free, small-world, locally-treelike and hierarchically-organized. Thus, the network approach turned out to describe complexity of seismicity in a peculiar manner.

There, however, remains an important problem, which is concerned with the cell size, i.e., the scale of coarse graining. In the previous investigations, it was nothing but a free parameter in the construction and no *a priori* rules were known for determining its value.

In this article, we address this problem by investigating the properties of the networks constructed from the real data taken from California, Japan and Iran. We observe that, remarkably, these three networks exhibit the universal scaling behavior. In particular, we find that the values of the exponent of the power-law connectivity distribution [8] and the clustering coefficient [9] of each network approach the respective universal values under the change of the cell size, depending on the number of events included in the data analyzed. This result turns out to offer a method of determining the cell size based on the invariance principle.

Quantities characterizing a network are usually dimensionless. In the case of an earthquake network, their values depend on the cell size, $L$. Therefore, it is natural to examine the following two dimensionless combinations:

$$l_3 = L / (L_{LAT} L_{LON} L_{DEP})^{1/3}, \qquad (1)$$

$$l_2 = L / (L_{LAT} L_{LON})^{1/2}, \qquad (2)$$

where $L_{LAT}$, $L_{LON}$ and $L_{DEP}$ are the dimensions of the whole geographical region under consideration in the directions of latitude, longitude and depth, respectively. Apparently, $l_3$ might seem more reasonable than $l_2$. However, in California, Japan and Iran, the earthquake networks are quite two-dimensional, since the majorities of events occur in the shallow regions there (see the explanation given below). We shall examine both of them.

As reported in Refs. [6,7], an earthquake network is scale-free and small-world. Accordingly, in what follows, we study the dependence of the quantities characterizing these properties on the cell size.

The seismic data we employ here are currently available at (i) California; http://www.data.scec.org, (ii) Japan; http://www.hinet.bosai.go.jp, (iii) Iran; http://irsc.ut.ac.ir/. The periods and the geographical regions covered are as follows: (i) between 00:25:8.58 on January 1, 1984 and 23:15:43.75 on December 31, 2006, 28.00°N–39.41°N latitude, 112.10°W–123.62°W longitude with the maximal depth 175.99 km, (ii) between 00:02:29.62 on June 3, 2002 and 23:54:36.21 on August 15, 2007, 17.96°N–49.31°N latitude, 120.12°E–156.05°E longitude with the maximal depth 681.00 km, (iii) between 03:08:11.10 on January 1, 2006 and 18:26:21.90 on December 31, 2008, 23.89°N–43.51°N latitude, 41.32°E–68.93°E longitude with the maximal depth 36.00 km, respectively. The total numbers of events in these periods are (i) 404106, (ii) 681547, (iii) 22845. As mentioned above, most of the events occurred in the shallow regions. In fact, 90 % of the events are shallower than (i) 13.86 km, (ii) 69.00 km, (iii) 26.60 km. The values of $(L_{\text{LAT}} L_{\text{LON}} L_{\text{DEP}})^{1/3}$ and $(L_{\text{LAT}} L_{\text{LON}})^{1/2}$ in eqs. (1) and (2) are respectively as follows: (i) 617.80 km and 1157.51 km, (ii) 1973.78 km and 3360.26 km, (iii) 566.25 km and 2245.73 km.

First, let us discuss the scale-free nature. The connectivity distribution of earthquake network is known to decay as a power law [6]

$$P(k) \sim \frac{1}{k^{\gamma}}. \tag{3}$$

One might naively expect from the procedure of network construction that the exponent, $\gamma$, should monotonically decrease as the cell size increases since the larger the cell size becomes, the more there appear hubs (i.e., vertices with large values of connectivity). However, unexpectedly, $\gamma$ turns out to approach a fixed value, remaining invariant, as the cell size becomes larger than a certain value, $l_*$.

We have analyzed for the scale-free earthquake networks the dependence of $\gamma$ on the cell size by making use of the method of maximum likelihood estimation [10]. The result is presented in Fig. 2. As one can see there, $\gamma$ *approaches a universal value*, ~ 1, as the cell size increases. Its behavior in Iran looks, however, different from those in California and Japan for the smaller cell size. In this respect, recall that the number of events contained in the Iranian data available for us is relatively small. Accordingly, we have cut and shortened the data from California and Japan in order to make their numbers of events adjusted to the Iranian one. The result is presented in Fig. 3. The difference of the global trends between California, Japan and Iran is seen to be diminished, showing an approximate scaling behavior.

Next, let us consider the clustering coefficient [9]. In this case, we have to reduce a full earthquake network to a simple network as schematically depicted in Fig. 1. Then, the reduced simple network is characterized by the symmetric adjacency matrix, $A$, satisfying the property: $(A)_{ij} = 1(0)$ if the $i$th and $j$th vertices are connected (unconnected). From it, the clustering coefficient, $C$, is defined by

$$C = \frac{1}{N} \sum_{i=1}^{N} c_i, \tag{4}$$

$$c_i = \frac{(A^3)_{ii}}{k_i(k_i-1)/2}, \tag{5}$$

where $N$ and $k_i$ are the total number of vertices contained in the network and the value of connectivity of the $i$th vertex, respectively. A reduced earthquake network is a small-world network [6,7], and its clustering coefficient is much larger than that of the corresponding random network (see Ref. [9]). One might expect that $C$ should increase as the cell size increases since the larger the cell size becomes, the easier triangle loops form [the $A^3$-structure in Eq. (5)]. However, again quite unexpectedly, $C$ also turns out to approach a fixed value, remaining invariant, as the cell size becomes larger than a certain value, $l_*$.

In Figs. 4 and 5, we present the dependence of $C$ on the cell size. In particular, Fig. 5 shows the result, in which the numbers of events in California, Japan and Iran are adjusted to each other, as in Fig. 3. A clear universal feature is observed there. Similarly to the case of $\gamma$, *C approaches a universal value,* ~ 0.85, for the large cell size. In addition, the difference of the global trends between California, Japan and Iran is seen to be reduced in Fig. 5, similarly to Fig. 3.

Thus, both $\gamma$ and $C$ exhibit the universal scaling properties. The (approximate) data collapses in $\gamma$ and $C$ in California, Japan and Iran, as shown in Figs. 3 and 5, enable one to determine the minimum cell size, $l_*$, which depends on the total number of events and above which these quantities become invariant. Very importantly, *both* of the universal values of $\gamma$ and $C$ yield a common value, $l_{2*} = 0.04 \sim 0.05$, as the minimum cell size in the present case of the numbers of events. (As recognized in Figs. 2-5, the scaling behaviors are better appreciated if $l_2$ is employed. This feature can be understood in terms of the fact that the majorities of events occur in the shallow regions, as mentioned earlier.) We confidently believe that the scaling behavior may become more manifest if the larger numbers of events are included in the analysis.

In conclusion, we have discovered remarkable scaling behaviors of the exponent of the power-law connectivity distribution, $\gamma$, and the clustering, $C$, in terms of the scale of coarse graining (i.e., the cell size) that is needed for constructing earthquake network. We have demonstrated by analyzing the earthquake networks in California, Japan and Iran that both $\gamma$ and $C$ reach the respective universal values at the cell size, $l_*$, generically depending on the number of events, and then remain invariant as the cell size further increases. In this way, we have found a method for determining the scale of coarse graining, $l_*$. It is our opinion that the scaling behavior discovered here may be inherent in seismicity of the earth.

*   *   *

SA was supported in part by a Grant-in-Aid for Scientific Research from the Japan Society for the Promotion of Science.

# Figure Caption

Fig. 1   Schematic description of earthquake network. (a) A full directed network, (b) the simple undirected network reduced from (a). The vertices with dotted lines indicate the initial and final events.

Fig. 2   Dependence of the exponent, $\gamma$, of the power-law connectivity distribution on the cell size (●: California, □: Japan, ×: Iran). (a) Plots of $\gamma$ with respect to $l_3$, (b) plots of $\gamma$ with respect to $l_2$. All quantities are dimensionless.

Fig. 3   Same analysis as in Fig. 2 with the numbers of events in California and Japan being adjusted to the Iranian data, 22845. (i) California; between 00:06:28.31 on January 1, 2005 and 13:52:11.55 on November 11, 2006, (ii) Japan; between 00:02:52.93 on June 1, 2007 and 05:43:19.30 on July 27, 2007. In this case, the values of $(L_{\text{LAT}} L_{\text{LON}} L_{\text{DEP}})^{1/3}$ and $(L_{\text{LAT}} L_{\text{LON}})^{1/2}$ are (i) 359.78 km and 1153.51 km, (ii) 1770.87 km and 3171.84 km, respectively. For the cell size larger than $l_{2*} = 0.04 \sim 0.05$, $\gamma$ approximately becomes invariant. All the quantities are dimensionless.

Fig. 4　Dependence of the clustering coefficient, $C$, on the cell size (●: California, □: Japan, ×: Iran). (a) Plots of $C$ with respect to $l_3$, (b) plots of $C$ with respect to $l_2$. All quantities are dimensionless.

Fig. 5　Same analysis as in Fig. 4 with the numbers of events in California and Japan being adjusted to the Iranian data, as in Fig. 3. For the cell size larger than $l_{2*} = 0.04 \sim 0.05$, $C$ approximately becomes invariant. All the quantities are dimensionless.

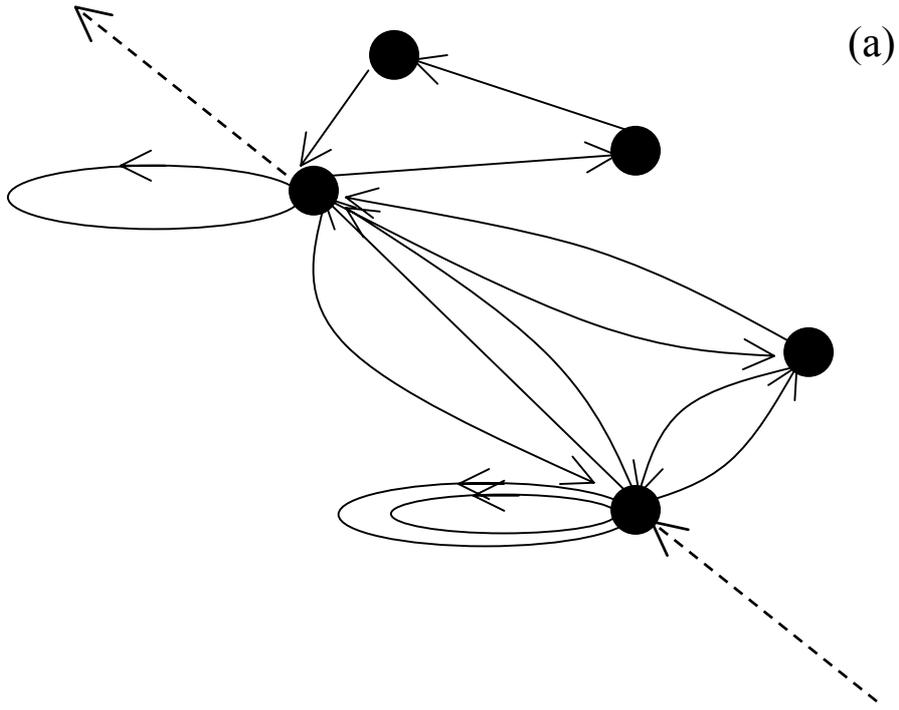

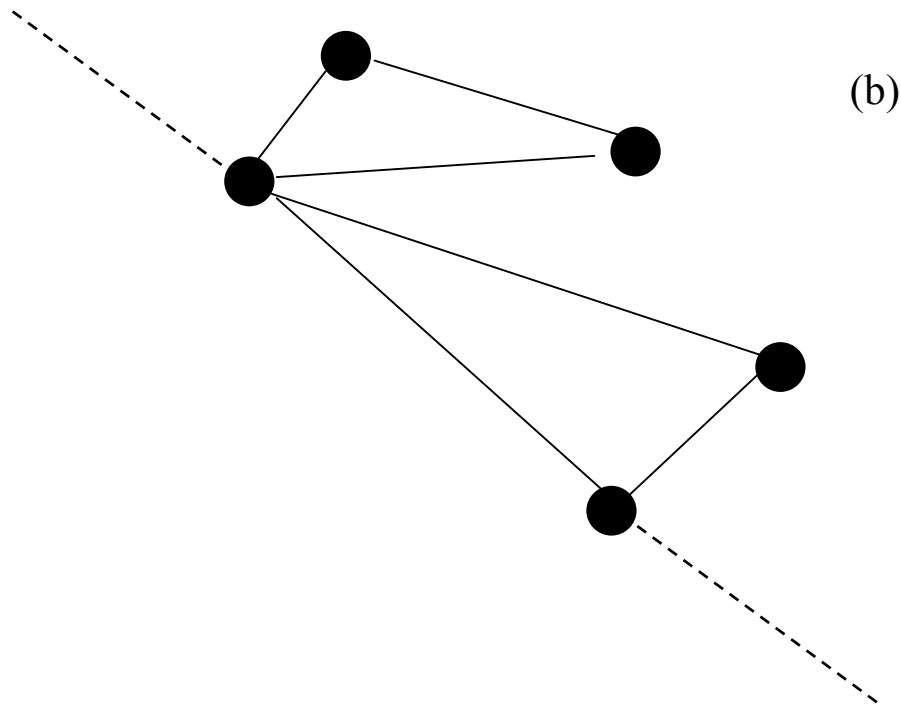

Fig.1

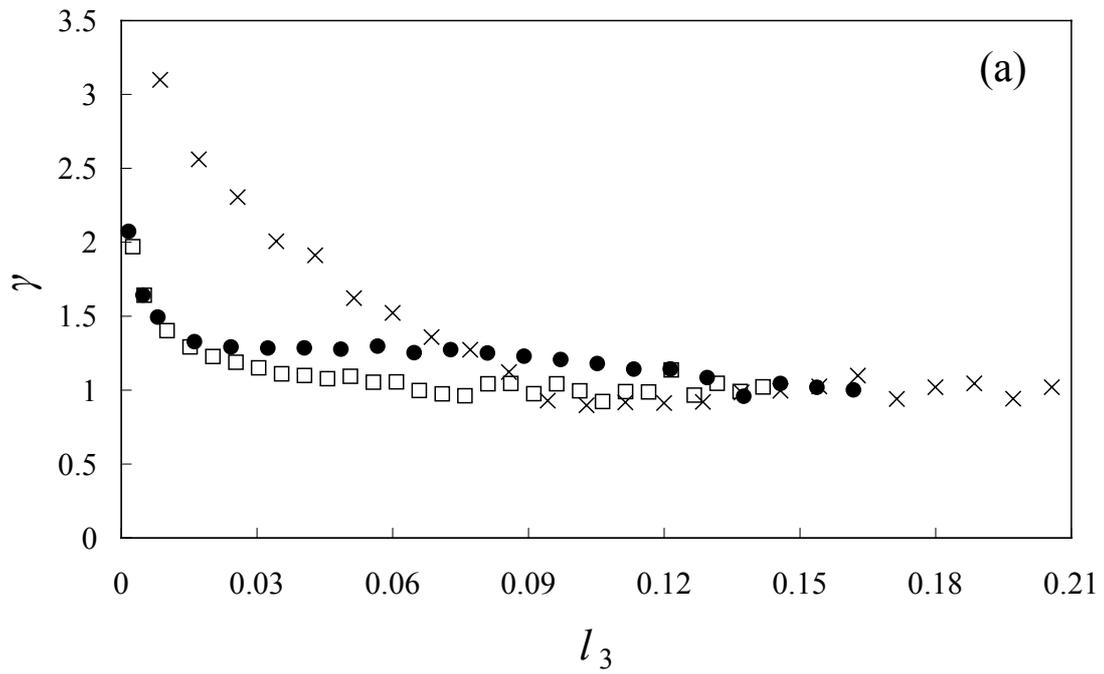

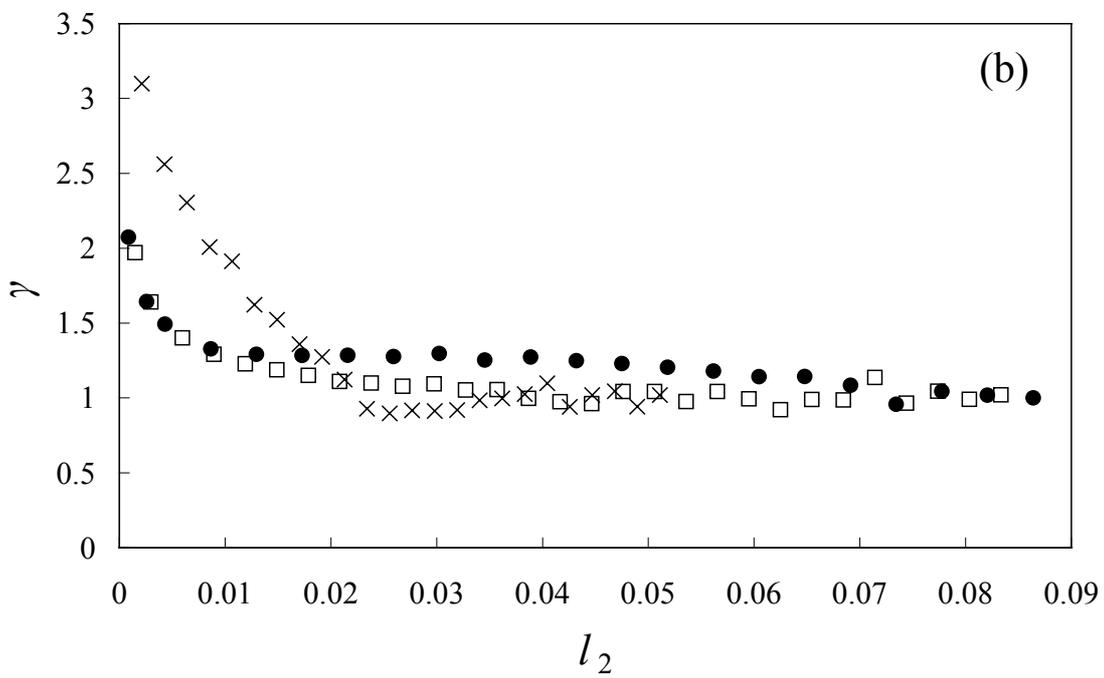

Fig.2

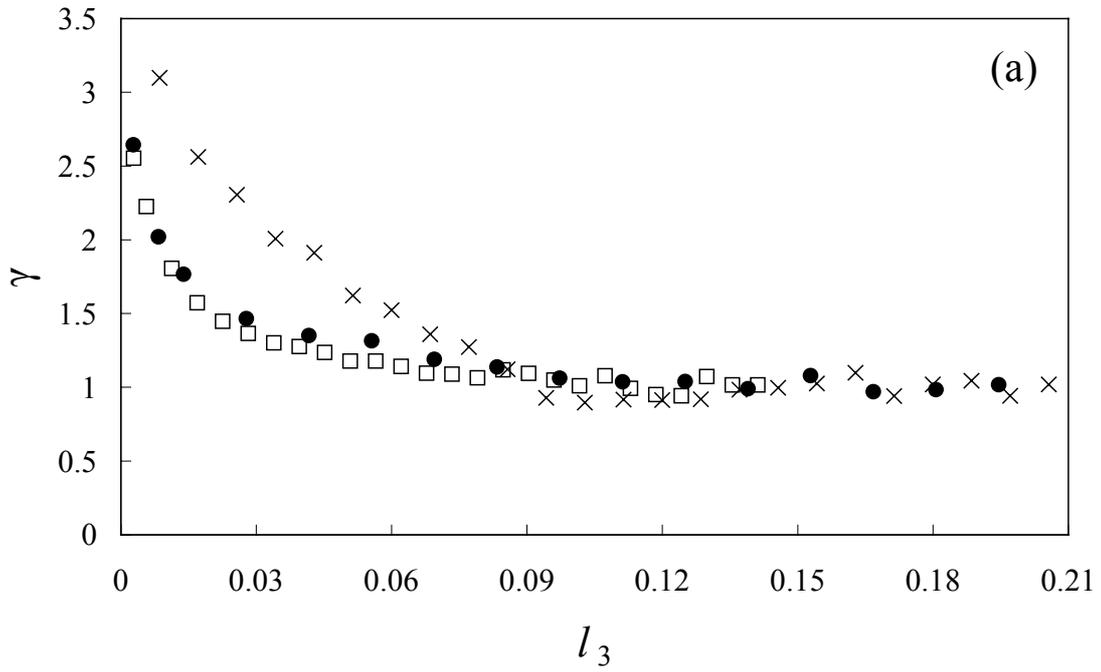

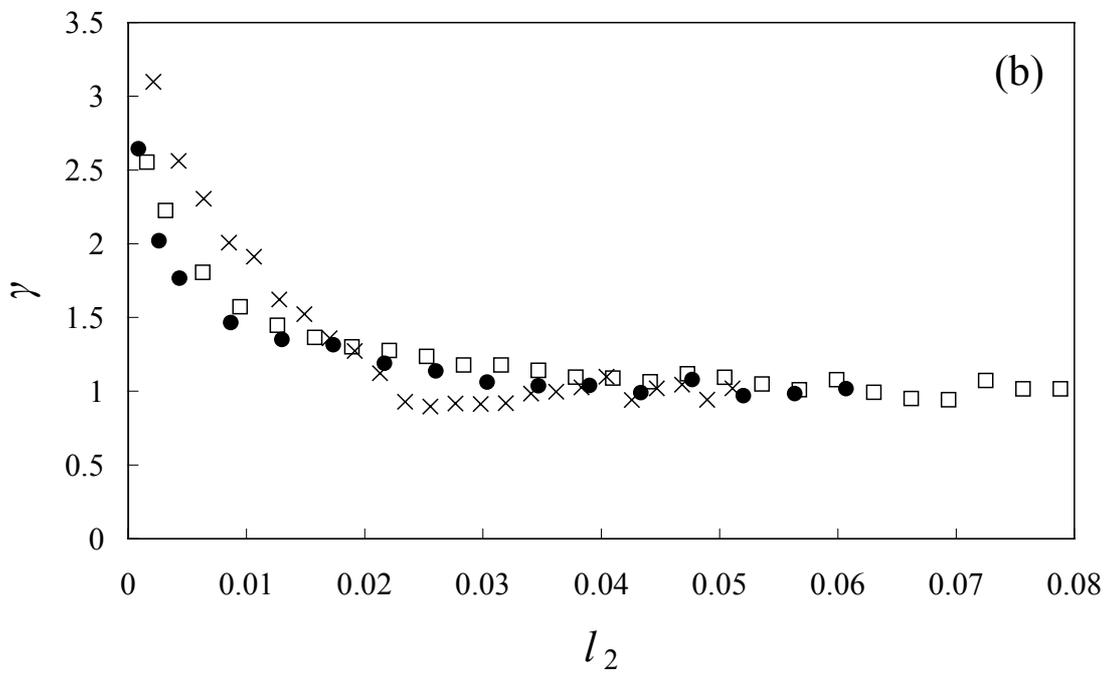

Fig.3

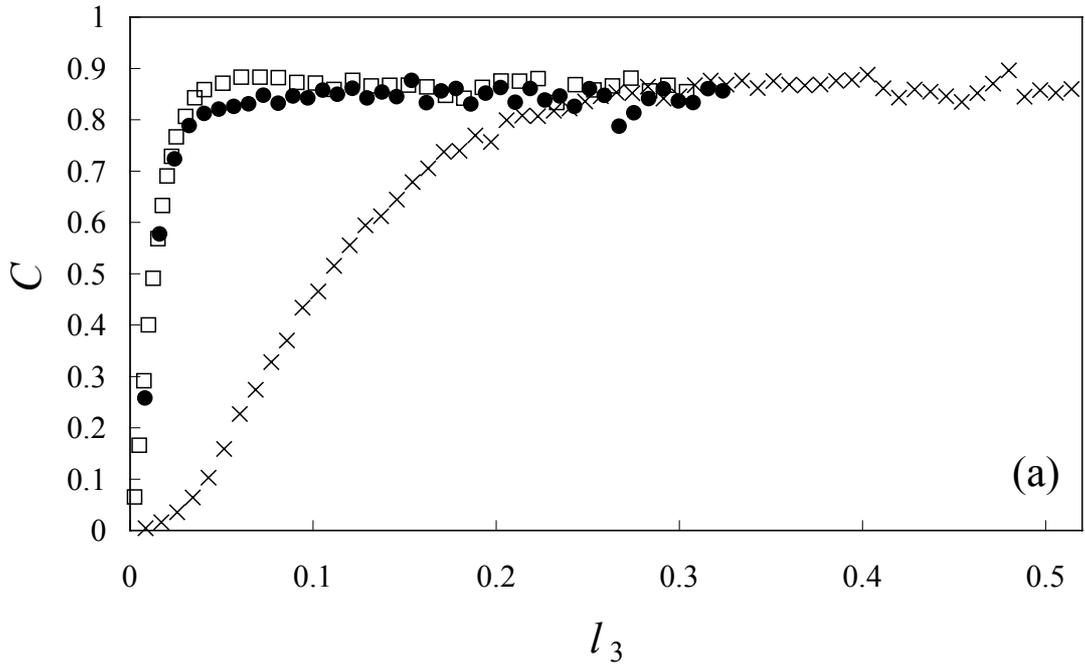

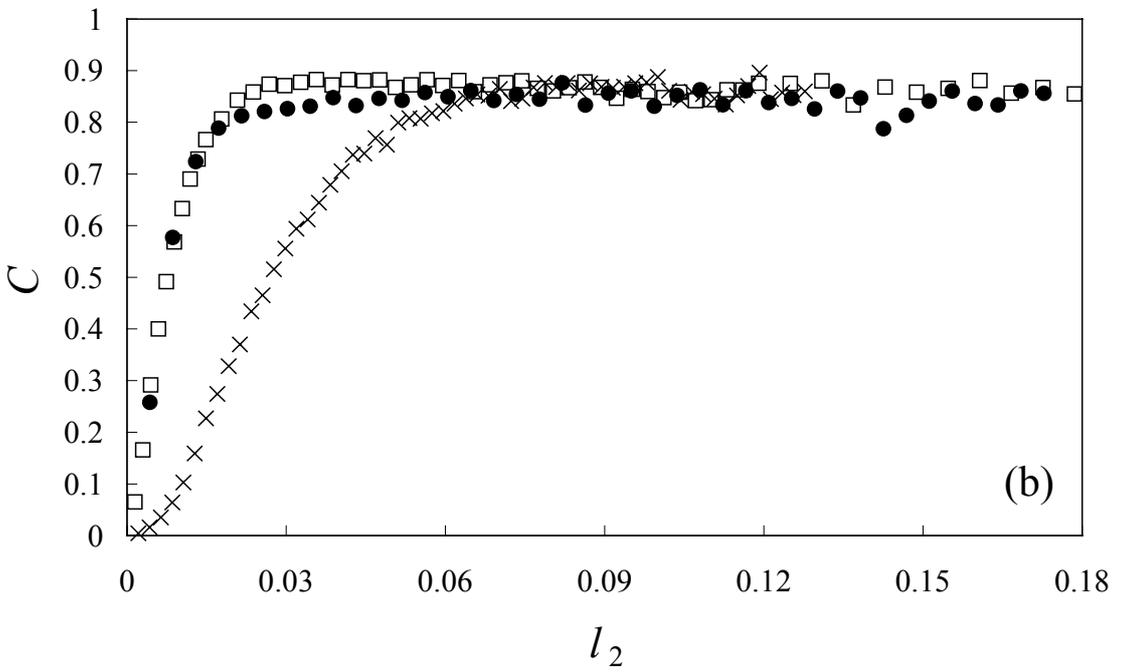

Fig.4

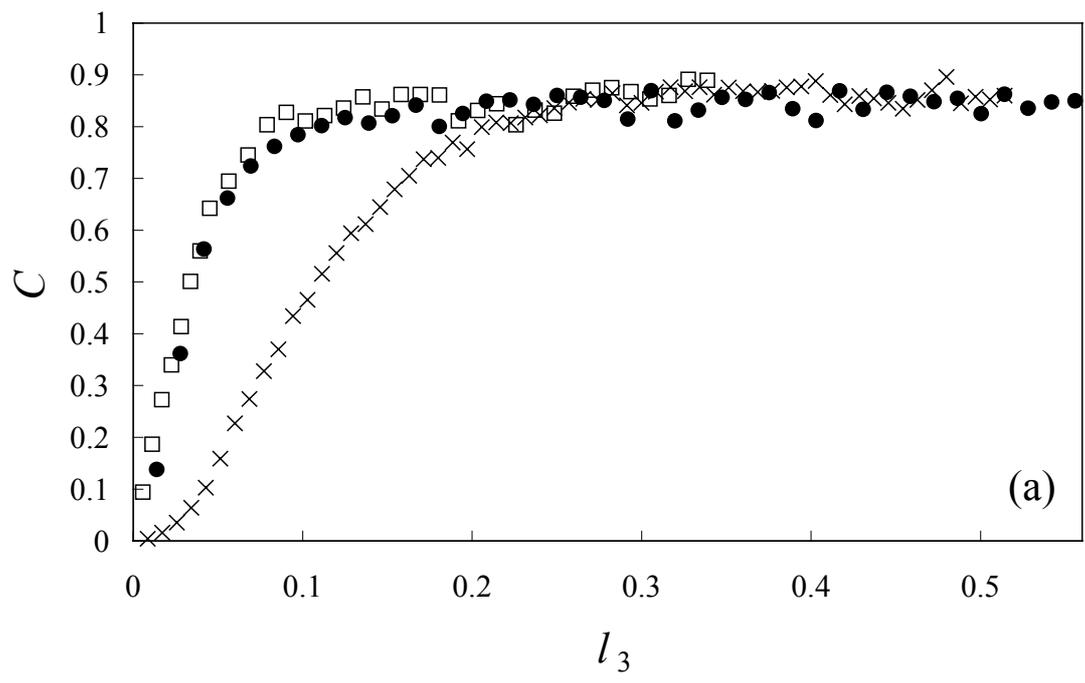

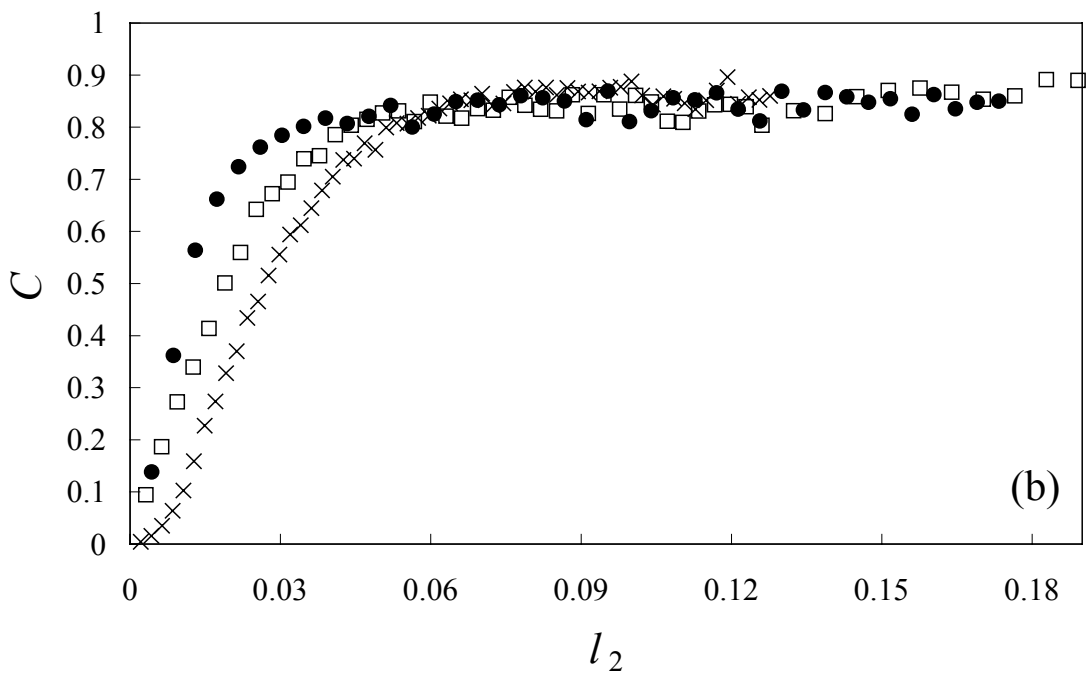

Fig.5